# Towards the Same Line of Liquid-liquid Phase Transition of Dense Hydrogen from Various Theoretical Predictions*

Binbin Lu(卢彬彬) [1†], Dongdong Kang(康冬冬) [1†], Dan Wang(王丹) [1],

Tianyu Gao(高天雨) [1], Jiayu Dai(戴佳钰) [1**]

[1]Department of Physics, National University of Defense Technology, Changsha 410073

*Supported by the National Natural Science Foundation of China under Grant Nos 11774429, 11874424, and U1830206, the Science Challenge Project under Grant No. TZ2016001, the National Key R&D Program of China under Grant No. 2017YFA0403200, the Science and Technology Project of Hunan Province under Grant No. 2017RS3038, and the Advanced Research Foundation of National University of Defense Technology under Grant No. JQ14-02-01.

**To whom correspondence should be addressed. Email: jydai@nudt.edu.cn

†These authors contributed equally to this work

%Tel. 0731-87001704    13787082112

**Abstract** For a long time, there have been huge discrepancies between different models and experiments concerning the liquid–liquid phase transition (LLPT) in dense hydrogen. In this work, we present the results of extensive calculations of the LLPT in dense hydrogen using the most expensive first-principle path-integral molecular dynamics simulations available. The nonlocal density functional rVV10 and hybrid functional PBE0 are used to improve the description of the electronic structure of hydrogen. Of all the density functional theory calculations available, we report the most consistent results through quantum Monte Carlo simulations and coupled electron-ion Monte Carlo simulations of the LLPT in dense hydrogen. The critical point of the first-order LLPT is estimated above 2000 K according to the equation of state. Moreover, the metallization pressure obtained from the jump of dc electrical conductivity almost coincides with the plateau of equation of state.

**PACS:**

Hydrogen at high densities displays rich phases and interesting phase transitions,[1],[2] which have long been a focus of both experiments and theoretical studies owing to its prominent role in condensed matter physics[3],[4] and planetary science.[5],[6] Despite the simplicity of the hydrogen atom, the structure and phase transition of the condensed phases of hydrogen at high densities still remain a great challenge today. Since atomic, solid, metallic hydrogen under high pressures was first predicted by Wigner and Huntington[7] in 1935, there have been intense efforts to pursue an accurate knowledge of the metallization and transition of hydrogen from its molecular-to-atomic phases. At low temperatures, three quantum molecular phases and a mixed molecular and atomic phase of dense hydrogen are observed below 315 GPa.[1]-[8] In addition, the phase diagram of solid hydrogen has recently been further enriched.[9],[10] The insulator-to-metal transition should occur when reaching at least 450 GPa according to Loubeyre's experimental estimate.[11] Recent measurements[12] also indicate that atomic metallic hydrogen may have been produced at a pressure of 495 GPa; however, this finding is yet to be confirmed.

At high temperatures above the melting line, the liquid–liquid phase transition (LLPT) of dense hydrogen at megabar pressures has been the subject of increasing interest in the last several decades. The location of the LLPT and the relationship between the atomic-to-molecular and insulator-to-metal transitions have become a central issue in high-pressure experiments in recent years.[13]-[21] Despite numerous experimental studies, different experiments have resulted in a controversial location of the LLPT of dense liquid hydrogen and deuterium.[16]-[18],[21] The dynamic compression reported by Knudson et al.[16] presents substantially higher metallization pressure measurements than static diamond anvil cell (DAC) measurements, and the transition pressure is nearly independent of temperature, which is inconsistent with static compression results.[17],[18] New dynamic measurements by Celliers et al.[21] show a broad pressure regime (approximately 100 GPa) between the onset of optical absorption and metallization and report that the signature of metallization in DAC measurements is correlated with the onset of absorption rather than that of metallization.

These experimental discrepancies have motivated a considerable number of theoretical studies on the LLPT of dense hydrogen.[22]-[33] There are a variety of theoretical methods with different approximation levels to address this issue. Of them, density functional theory (DFT)-based first-principle molecular dynamic (FPMD) simulations and quantum Monte Carlo (QMC) simulations have become the most popular approaches to solve many-body quantum systems. In the framework of FPMD, electrons are quantum-mechanically described in the Kohn–Sham scheme of DFT and ions propagate on the electron-produced energy surface in accordance with Newton's equation of motion. Although FPMD simulations have been extensively applied to studying material properties, the accuracy of this method in predicting the location of the LLPT is still limited by the following two approximations: one is the local or semi-local density functionals employed in DFT calculations, e.g., the Perdew–Burke–Ernzerhof (PBE) functional,[34] while the other is the neglect of the nuclear quantum effects (NQEs) of protons in computer simulations.[35]-[37] The PBE functional typically underestimates the band gap by 1–2 eV[38] in hydrogen, resulting in a lower metallization pressure. Calculations with nonlocal functionals, i.e., vdW-DF1[39] and vdW-DF2,[40] predict much higher transition pressures between the insulating molecular fluid and the metallic atomic fluid.[28],[31],[32] As reported by Li et al.,[41] the rVV10[42],[43] and vdW-DF1 nonlocal functionals exhibit better agreement with multi-shock compression measurements of hydrogen–helium mixtures than other functionals. While dispersion interactions affect the accurate location of the LLPT of hydrogen, the self-interaction error also plays a non-negligible role in improving the description of the electronic structure of high-pressure hydrogen, as reported by Morales et al.[28] Moreover, high-precision experiments show that no single exchange-correlation functional describes both the onset of dissociation and the maximum compression along the Hugoniot[20] well. Therefore, a satisfactory density functional is key to obtain an accurate description of the structure and phase transition of dense hydrogen. In addition, due to the lowest mass element, the NQEs of protons strongly influence the structure and dynamic properties of dense hydrogen,[35]-[37] thereby affecting the dissociation process of liquid hydrogen. Therefore, the classical treatment of protons in previous FPMD simulations would certainly result in non-negligible errors for the LLPT in dense hydrogen.[28] The QMC approach,[44] unlike the density-based DFT, is a wave-function-based method. The QMC simulations are thought to be more accurate than DFT methods, although they are computationally more expensive than DFT methods. With the continuing improvements of specific implementations, the quality of the variational wave function, and the finite-size effects errors, QMC calculations are expected to provide benchmark results for the dissociation process and metallization transition of hydrogen at megabar pressures.[27],[28],[30]-[32]

Based on the theoretical approaches mentioned above, the LLPT in liquid hydrogen has been greatly explored. The existence of a first-order LLPT in dense liquid hydrogen has been indicated by both FPMD and QMC calculations.[26]-[32] The critical point, separating the continuous crossover regime and the first-order transition regime, is predicted to exist at temperatures greater than 10,000 K with chemical models[45],[46] but decreases to 1500 K based on FPMD simulations with the PBE functional.[26] Recent coupled electron-ion Monte Carlo (CEIMC) simulations have estimated the critical point of the LLPT to be at temperatures and pressures near 2000 K and 120 GPa,[27] respectively. Below the critical point, the first-order LLPT in dense hydrogen is characterized by the equation of state (EOS), pair-correlation function (PCF), and electrical conductivity in previous studies, where the PCF and electrical conductivity exhibit a sharp signature at the strong first-order transition, whereas the EOS shows a plateau at the transition pressure.

In this study, we present the results of extensive calculations of the LLPT in dense liquid hydrogen using first-principle PIMD[47],[48] simulations with a recently proposed van der Waals density functional rVV10 to account for the dispersion interactions of electrons, since it is thought to be a promising functional to provide a better description of the electronic dispersion interactions compared to previous candidates and suitable from gas to solid phases.[42],[43] vdW-DF1 functional is also used since it uses the same exchange functional as rVV10 and considered to be a good choice for dense hydrogen.[21] A robust hybrid density functional PBE0 using Hartree–Fock exchange with a truncated Coulomb operator[49] is also used to obtain the accurate description of the electronic structure. As a result, we obtain the accurate location of the LLPT in dense liquid hydrogen that has the best agreement with the QMC and CEIMC results relative to other DFT-based calculations.

PIMD simulations were performed using the generalized Langevin dynamics implemented i-PI code,[50][51] which was driven by DFT calculations with Quickstep package.[52] At least 10000 steps with the 0.5-fs time step were run in MD simulations, while more than 10000 steps with the 0.2-fs time step were run in PIMD simulations. 16 beads were used to sample the imaginary-time path integral at each temperature. In DFT calculations, the bands were occupied by electrons according to the Fermi–Dirac distribution function. Wave functions were expanded in a DZVP Gaussian basis set, where a Gaussian was mapped onto the finest grid and the grid of the energy cutoff of 500 Ry and 50 Ry to achieve good convergence. In order to accurately predict the location of a phase transition, a big supercell including 256-atom with $\Gamma$ point for the representation of the Brillouin zone is employed, therefore, the finite-size effect errors can be neglected (see the Supplementary Material). The electronic density of state (DOS) was calculated using the Quantum-ESPRESSO package.[53] The electrical conductivity was calculated using the Kubo-Greenwood formulation based KGEC package.[54] Ten uncorrelated ionic configurations along the trajectory were sampled for the DOS and electrical conductivity calculations.

We simulated the equilibrium states of liquid hydrogen over a wide range of densities along four isotherms: 600 K, 1000 K, 1500 K, and 2000 K. The results of the location of the LLPT from PIMD simulations with the rVV10 and vdW-DF1 functionals and the corrections with the PBE0 hybrid functional are shown in Fig. 1.

We can see that the trend in the phase boundary from our quantum simulations is in good agreement with the static compression results, even though the predicted transition pressures appear to be shifted to higher pressures by approximately 20 GPa compared to the experiments.[17],[18]

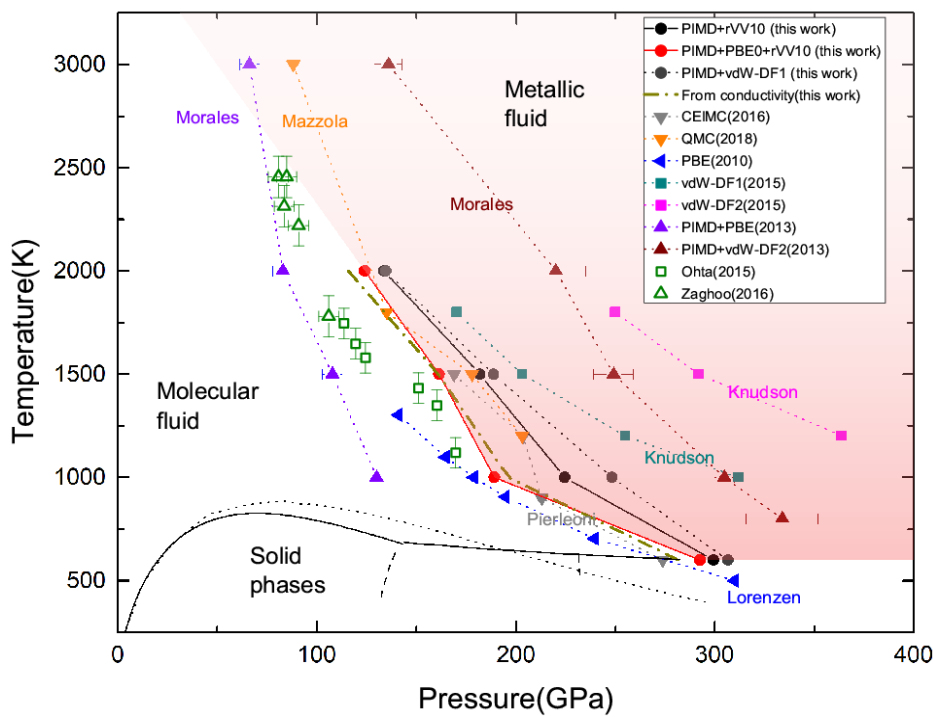

Fig. 1. Phase diagram of dense hydrogen with experimental data and theoretical predictions from this study as well as those of QMC and CEIMC simulations. Phase transition points from our PIMD simulations with the rVV10 (black circles) and vdW-DF1 (gray circles) functionals, as well as the correction with hybrid functional PBE0 (red circles), are presented. The static compression results are displayed using the green upward triangles (ref. [17]) and green (ref. [18]) squares. The blue leftward triangles, green squares, and pink squares represent the DFT calculations with the PBE,[26] vdW-DF1, and vdW-DF2 functional,[16] respectively. The orange downward triangles correspond to the QMC-based molecular dynamic simulations,[31] while the gray downward triangles refer to the CEIMC predictions of the LLPT.[32] The purple upward triangles and brown upward triangles refer to PIMD calculations with the PBE and vdW-DF2, respectively.[28] The insulator-to-metal transition line (dot-dashed line) obtained from this study is also presented.

As shown in Fig. 1, the phase boundary between the molecular liquid and the atomic liquid obtained from different theoretical approaches is distributed over a fairly wide range of pressures (~150 GPa). Of all the DFT-based calculations, the prediction of the transition pressure using PIMD simulations with vdW-DF1 is substantially larger than those of the CEIMC[32] and QMC[31] results, especially in the high pressure range, while the transition point using the PBE functional is shifted to lower pressure.[26] When considering the NQEs, PIMD calculations with PBE predict a much lower transition pressure than measurements and other calculations[28]. Conversely, PIMD calculations with vdW-DF2 overestimate the transition pressure between the two liquid hydrogen states[28]. Therefore, these nonlocal density functionals cannot provide a satisfactory description of the LLPT of dense liquid hydrogen. We can see that the location of the LLPT predicted using our PIMD calculations with the rVV10 functional is in good agreement with the CEIMC and QMC results. Celliers et al.[21] suggested that the vdW-DF1 functional is currently the best choice for the insulator–metal transition for dense hydrogen. However, in our calculations, vdW-DF1 somewhat overestimates the transition pressures at high pressures while it performs as good as rVV10 at low pressures. This is because both the rVV10 and vdW-DF1 functionals have the same exchange functional part of the van der Waals interactions and the different treatments of the correlation functionals predict different transition points for dense hydrogen. The PBE0 corrections from 27 GPa at 600 K to 8 GPa at 2000 K result in the phase boundary being closer to the CEIMC and QMC predictions. The effects of the PBE0 corrections of 6% and 1% for the pressure and internal energy, respectively, are similar to the QMC calculations relative to the PBE calculations, where the QMC simulation predicts a pressure that is ~5% smaller than the PBE results.

Here, the LLPT of dense liquid hydrogen is characterized by the EOS and PCF along four isotherms (see Fig. 2). In particular, the transition pressures are determined by the discontinuity in the curves of pressure versus density denoted by the Wigner–Seitz radius $r_s$. There is always a plateau when a first-order transition occurs with increasing pressure. In PIMD simulations with rVV10 and PBE0 functionals, this plateau is clearly displayed at all the temperatures in this study, while the plateau in the vicinity of the transition point is gradually disappeared with increasing temperature without the inclusion of the PBE0 correction. That indicates a continuous crossover between the two liquid states. Therefore, in our PIMD simulations with rVV10 and PBE0 functionals the critical point of the first-order LLPT of dense liquid hydrogen at least above 2000 K, which is in consistent with the recent prediction,[55] and higher than predicted by PBE[26] and CEIMC[27] calculations. In addition, we can see from Figs. 2b and 2c that the nonlocal density functional rVV10 predicts transition pressures that are more consistent with the QMC and CEIMC results than vdW-DF1, indicating the superiority of the rVV10 functional for calculating the LLPT of dense liquid hydrogen.

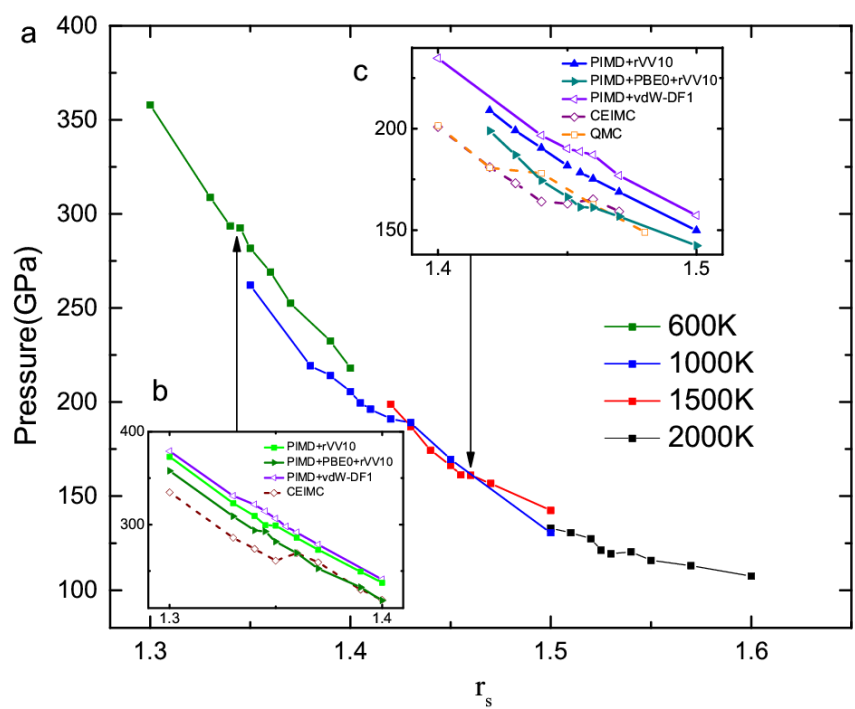

Fig. 2. EOS of dense liquid hydrogen. (a) our PIMD calculations with the rVV10 and PBE0 functionals at temperatures of 600 K, 1000 K, 1500 K, and 2000 K. (b) Comparisons of the EOS of the DFT-based PIMD calculations and the CEIMC[32] results (purple diamonds) at 600 K. The green squares, green rightward triangles, and violet leftward triangles represent the results of the PIMD simulations with rVV10, PBE0+rVV10, and vdW-DF1, respectively. (c) comparisons of the EOS of the DFT-based PIMD calculations and the QMC[31] (yellow squares) and CEIMC[32] results (purple diamonds) at 1500 K.

Molecular dissociation is characterized by the PCF with density. In Fig. 3a, the height of the first peak of the PCF from the PIMD calculations at 600 K is gradually reduced with increasing density, even though it has an obvious decrease at the transition plateau ($r_s$=1.345). Nevertheless, the molecular peak does not disappear and becomes a shoulder at high density ($r_s$ = 1.30). The PCF exhibits similar behaviors at temperatures from 600 K to 2000 K, indicating that the hydrogen molecules undergo a gradual dissociation process with increasing density. In comparison with the CEIMC calculations,[33] we note from Fig. 3b that PCF in our PIMD calculations with the rVV10 functional is less structured, indicating a higher molecular dissociation degree than CEIMC results.

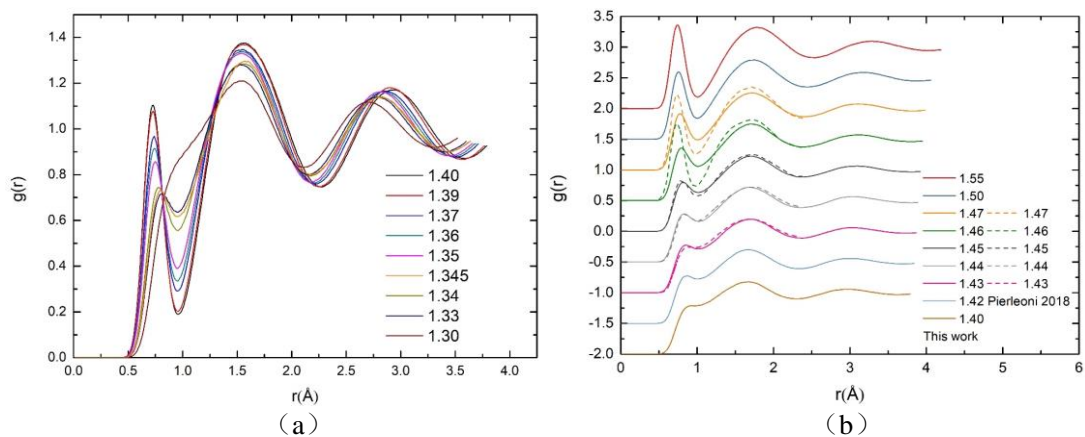

Fig. 3. PCFs of dense liquid hydrogen. (a) comparisons of the PCFs for different densities at 600 K obtained from calculations of the PIMD with rVV10 (b) comparisons of PCFs of our PIMD calculations with rVV10 (solid lines) to the CEIMC[33] results (dashed lines) at 1500 K.

The insulator-to-metal transition of dense hydrogen is characterized by the sudden jump of dc electrical conductivity with increasing pressures. Fig. 4 shows the trend of electrical conductivity with increasing pressures in our calculations with differing functionals. There is remarkable rapid increase when the pressure is increased from 600 K to 2000 K. Meanwhile, the pressure corresponding to the jump of electrical conductivity is quite different with differing functionals. Here we use the minimum metallic conductivity of 2000 $\Omega^{-1}cm^{-1}$ as the criteria to determine the insulator-to-metal transition point.[55] In fact, as shown in Fig. 4, the pressures corresponding to the minimum metallic conductivity are in the pressure range of electrical conductivity jump. At 1500 K, the results from recent CEIMC simulations [55] with HSE functional [56] used in electrical conductivity calculations is presented for comparisons. The electrical conductivities obtained from the two approaches exhibit similar behavior with increasing pressure. Although the electrical conductivity in this study is slightly higher than the CEIMC calculation at low pressures, [55] the insulator-to-metal transition point only has a small difference of about 10 GPa.

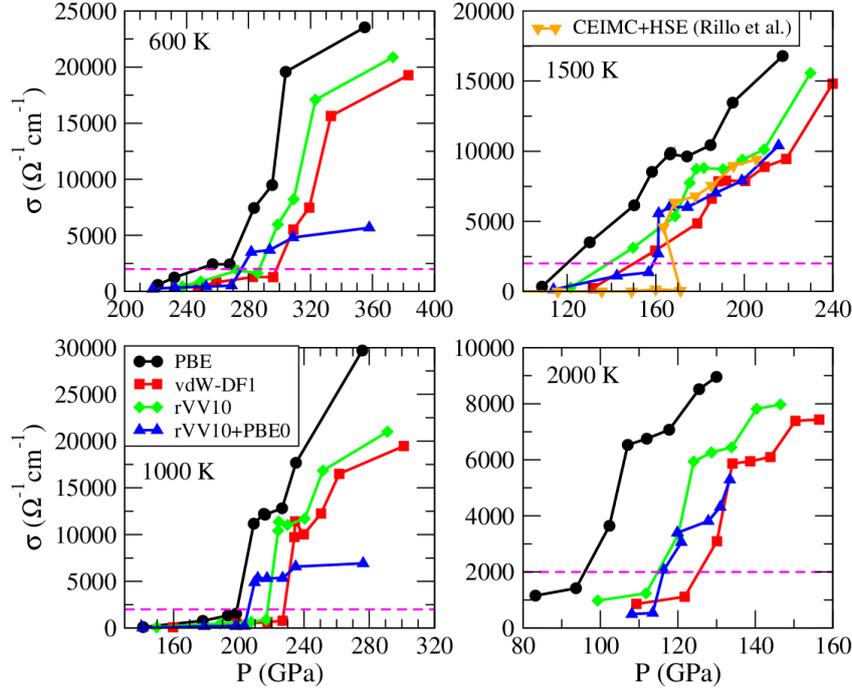

Fig. 4. Electrical conductivity of hydrogen as a function of pressure at 600 K, 1000 K, 1500 K and 2000 K. The results with the PBE, vdW-DF1, rVV10, and rVV10+PBE0 functionals are presented for comparisons. The horizontal dashed lines represent the minimum metallic conductivity of 2000 $\Omega^{-1}\text{cm}^{-1}$. The results obtained from CEIMC simulations [55] and HSE functional at 1500 K are presented.

The insulator-to-metal transition line obtained from our PIMD simulations with PBE0 functional is shown in Fig. 1. We find that the metallization is almost accompanied by the discontinuity of EOS. Therefore, the metallization pressure is the criteria to determine the LLPT of dense liquid hydrogen. Here we should note that the infrared optical measurement probes the onset of optical absorption and always underestimate the metallization pressure because of the high photon energy.

Terahertz-frequency optical measurement can obtain the accurate dc electrical conductivity and should be applied for the metallization probe of dense liquid hydrogen.[57]

The metallization process can also be characterized by DOS. Fig. 5 shows the DOS of dense hydrogen at 1000 K obtained with differing functionals. Metallization occurs at a density of $r_s = 1.42$ according to the PIMD calculations with rVV10. When considering the PBE0 correction, the DOS of hydrogen still has a band gap at $r_s = 1.42$ and the metallization density shifts to a higher density of $r_s = 1.41$, indicating that the PBE0 correction causes the metallization density to be higher. However, as mentioned above (see Fig. 1), the PBE0 correction lowers the LLPT pressures. That is because the PBE0 has significantly corrections to pressure (see Supplementary), i.e., greatly lowering the phase transition pressure.

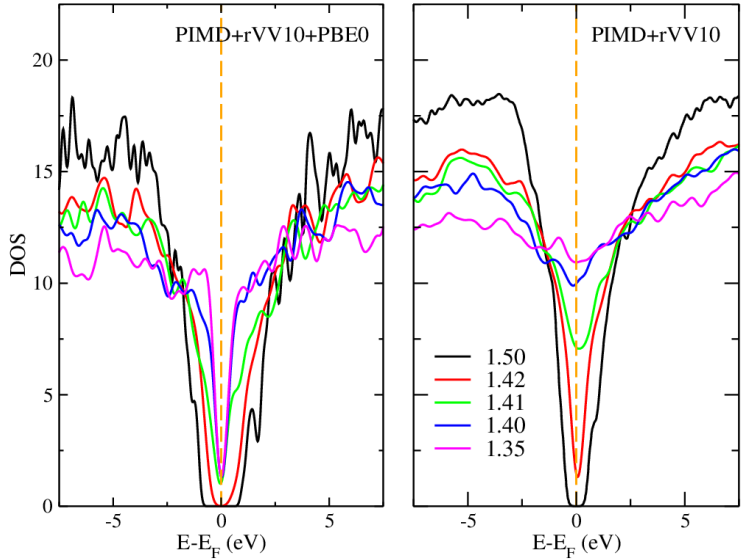

Fig. 5. DOS of dense hydrogen with different density functionals at 1000 K (a) DOS obtained from PIMD calculations with the rVV10 and PBE0 functionals (b) DOS obtained from PIMD calculations with the rVV10 functional.

The hybrid functional PBE0 corrections to EOS and electrical conductivity result from the accurate description of electronic structure of dense liquid hydrogen. On one hand, PBE0 functional largely improve the self-energy error of the semi-local approximation of exchange-correlation functional, such as PBE. On the other hand, electrons exhibit the localized state character at the top of valence band from the local DOS (see Supplementary), and PBE0 functional could provide a proper description of such localized electrons.[58] Therefore, when taking into account the hybrid functional corrections in our quantum simulations, the LLPT obtained in this study becomes closer to both the static experiment measurement and QMC and CEIMC calculations.

In conclusion, we investigated the LLPT of dense liquid hydrogen at megabar pressures and at temperatures below 2000 K using first-principle PIMD simulations with the nonlocal density functionals rVV10 and vdW-DF1 and the hybrid density functional PBE0, which were used to improve the description of the electronic structure. First, we obtained an accurate location of the LLPT in dense liquid hydrogen using state-of-the-art simulations; compared to other DFT-based calculations, our simulation is in best agreement with the QMC and CEIMC results. We find that the rVV10 functional is the best choice for the LLPT of dense liquid hydrogen and that the critical point of the first-order LLPT is above 2000 K. The metallization pressure obtained from the jump of dc electrical conductivity almost coincides with the plateau of EOS. Second, the molecular dissociation of hydrogen occurs over a fairly wide range of pressure, even though there is an obvious decrease in the number of molecules in the vicinity of the transition point. Third, when the electronic structure is described more accurately using the PBE0 correction, we find that the PBE0 correction results in a lowered transition pressure for the LLPT and shifts the metallization to higher densities. Finally, to confirm the theoretical predictions of the structure and dynamical properties of dense hydrogen, more high-precision measurement techniques need to be developed.